\documentstyle{article}
\begin{document}
\centerline{\bf Immune Network: An Example of Complex Adaptive Systems} 

\vspace{1cm}

\centerline{\bf Debashish Chowdhury} 
\centerline{Physics Department}
\centerline{Indian Institute of Technology}
\centerline{Kanpur 208016, U.P.}
\centerline{India}

\vspace{1.5cm}

{\bf Abstract:} 
The phenomenon of immunological memory has been known for a long time. 
But, the underlying mechanism is poorly understood. According to the 
theory of clonal selection the response to a specific invading antigen 
(e.g., bacteria) is offered by a specific clone of the cells. Some of 
the lymphocytes activated during the primary response remain dormant 
and keep circulating in the immune system for a long time carrying the 
memory of the encounter and, therefore, these long-lived cells are 
called memory cells. Proponents of the alternative network theory 
maintain that the immune response is offered by a "network" of clones 
in a collective manner. In recent years several possible scenarios of 
the "structure" and function of the immune network have been considered. 
We have developed mathematical models for describing the population 
dynamics of the immunocompetent cells in a unified manner. We have 
incorporated intra-clonal as well as inter-clonal interactions in a 
discrete formulation and also studied a continuum version of this 
model. 

\vspace{2cm}

\section{Introduction:}

The latin word "immunitas" is related to the concept of exemption from 
a service or duty or from civil laws (e.g., "diplomatic immunity" of an 
ambassador of one country in another). It has been known for more than two 
thousand years \cite{silver89} that individuals who recover from a 
disease become "immune" to it; this is the phenomenon of "acquired immunity". 
The scientific investigation of immunology, however, began much later 
when Jenner utilized this phenomenon of "acquired immunity" to develope 
a vaccine against small pox. The first breakthrough in understanding the 
mechanism of this remarkable phenomenon was made by Louis Pasteur in 1880. 
Over the last hundred years we have collected an enormous amount of information 
on the "hardware" of the immune system (e.g., the molecules and cells 
involved) \cite{albert89, darnel90, jaret86} but we understand very little 
about the "software" that runs it, i.e., the principles governing various 
immunological processes. 

Theoretical immunology \cite{perweis, chow92}deals with the mathematical 
modelling of immunological processes at various levels, e.g., molecular level, 
cellular level and the level of cell populations. One of the major aims of 
theoretical immunology \cite{perel88, marchuk83} is to predict "macroscopic" 
properties of the immune system from the properties and interactions among 
its elementary "microscopic" constituents; this problem is similar to those  
usually studied by physicists using the techniques of statistical mechanics. 
Theoretical immunologists develop mathematical models to understand how the 
immune system evolves over long time scales, how its size and inter-cellular 
interactions vary with time, how these interactions govern the dynamics of 
the populations of various types of cells during an immune response to a 
specific antigen and how it "learns" adaptively about new antigens (i.e., 
acquires new knowledge) and how it retains the newly acquired knowledge in 
its "memory" and, finally, how it retrieves information from its memory. 
Several mathematical models have been developed so far to capture 
the known immunological phenomena as well as to predict new ones. In this 
chapter we summarize some of the modern approaches to the mathematical 
modelling of the immune system and illustrate these with specific examples. 
We hope that some of the modelling strategy developed for the immune system 
may find applications in designing artificial immune systems. 

\section{A brief summary of experimental phenomena to be modelled theoretically:}

Millions of different varieties of lymphocytes are known to be produced 
by the immune system. However, according to the {\it clonal selection theory}, 
only a specific type can respond to a specific antigen. This is in sharp 
contrast to the non-antigen-specific response offered by the macrophages 
to the antigens. The body seemingly anticipates all the types of antigens 
it may encounter in the future and prepares accordingly by producing a large 
variety of lymphocytes.  For an antigen-specific response the antigen must be, 
first of all, properly recognized by the specific lymphocytes.  Different 
types of lymphocytes identify the antigens in different manners.  Following 
the recognition of the antigen, a specific type of lymphocyte, which fit best 
with the antigen, proliferates rapidly through cell division into a clone (a 
population of genetically identical cells). The corresponding process is 
called clonal selection because the antigen selects which lymphocytes 
must develop into a clone \cite{burnet57}. 

There are several alternative and complimentary routes of immune response. 
In a humoral immune response a specific type of B-cell proliferates and the 
terminal differentiation of a fraction of this B-cell population leads to 
plasma cells. These produce antibodies, which react with the antigen and 
eventually lead to the elimination of the antigen from the host system. 
The remaining fraction of the proliferating B-cells become dormant and keep 
circulating in the bloodstream carrying a memory of the encounter with 
the antigen; the latter variety of the long-lived B-cells are called 
memory B-cells. In the cell-mediated immune response a specific type of 
T-cell becomes cytotoxic and kills the antigen directly. Memory of the 
encounter with the specific antigen is thereafter carried by the corresponding 
long-lived memory T-cells. The helper T-cells play very crucial roles of 
regulating the immune response in both routes to immunity. The host 
carrying the memory cells is said to have acquired immunity against the 
specific antigen because the presence of the memory cells leads to a quicker 
and stronger secondary immune response when the host is stimulated again 
with the same antigen. In fact, this is the basic principle of vaccination.  
The humoral and cell-mediated routes to immunity are illustrated schematically 
in fig.1. 

Normally, the immune system can distinguish the cells and tissues of the 
host ("self") from the foreign invaders ("non-self"). A normal immune 
response (NIR) follows when the population of a foreign antigen in the 
body exceeds a tolerance level \cite{reni90}. However, under special 
circumstances, the immune system mistakenly identifies a part of the host 
as a "foreign"  substance because of some "error" \cite{cohen88}. 
Then, the immune response that follows against the host is 
called an auto-immune response (AIR) and such a response can lead to a 
auto-immune disease. 

The human immunodeficiency virus (HIV) is an exceptional invader in the 
sense that, unlike other foreign antigens, it destroys the helper T-cells 
of the immune system which are known to be crucial for almost all types 
of immune response. Therefore, during the late stages of HIV infection the 
patient's immune system becomes disabled; this is the acquired immune 
deficiency syndrome (AIDS). Such a patient ultimately succumbs to a secondary 
infectious agent, raher than to the HIV itself \cite{weber}. 

\section{Clonal Selection and Its Mathematical Modelling:} 

The equations describing the population dynamics of the cells involved 
in immune response to a specific antigen can be formulated in two different 
ways. In the discrete approach, the population of each type of cells is 
modelled by a discrete variable which can take one of only two allowed 
values: $0$ and $1$ corresponding to low and high populations, respectively. 
In this approach, the dynamical equations are written as maps in discrete 
time. Moreover, the interactions between the various pairs of different 
cell types are also restricted to have only a few discrete set of allowed 
values. On the other hand, in the continuum approach, the populations of 
the cells, the interactions between the different cell types as well as 
time are assumed to be real variables, which can vary continuously; the 
population dynamics of the cells are now given by a set of differential 
equations. 

The continuum approach was followed first by Bell \cite{bell70, bell71}, 
and subsequently by many other investigators, for developing mathematical 
models of clonal selection. But, since quite often the models are 
"underdetermined" by the available experimental data, i.e., more 
than one model can account for the known experimental facts, some authors 
have, in the recent years, advocated the use of a discrete language 
as a first step towards the formulation of the quantitative theories 
\cite{atlan89}. The advantage of a discrete language arises from the fact 
that the range of allowed values of the variables and the parameters is 
so narrow that one does not need to adjust too many free parameters 
to reproduce experimentally known facts. The discrete theories 
can satisfactorily account for the qualitative features of immune response.  
The discrete models are not intended to be a substitute for the more 
realistic continuum description; one can construct the continuum counterpart 
of the discrete model following well known mathematical prescriptions.  
The relative advantages and disadvantages of these two approaches have also 
been discussed in detail in the literature.

\subsection{Discrete Models of Clonal Selection}

The discrete variable that describes the populations of the cells in the 
discrete models is sometimes called an "automaton" and a system consisting 
of such mutually-interacting automata are referred to as "cellular automata" 
(CA). The concept of CA was introduced by Von Neumann in the context of 
theories of evolution and, subsequently, analyzed in more detail by Wolfram 
\cite{wolf86}. The CA are known to exhibit a rich variety of spatio-temporal 
patterns depending on their rules of evolution. These have found practical 
applications in modelling, for example, fluid flow, etc. 

Discrete models of cell population dynamics of the immune system can be 
formulated in terms of either "threshold automata" or "Boolean automata" 
\cite{weisbook91}. 

\noindent $\bullet$ {\bf Threshold automata:} The population of the $i$-th 
type of cells is denoted by the symbol $S_i$, where each $S_i$ can take 
only two possible values: $S_i = 0$ corresponding to a low population and 
$S_i = 1$ corresponding to a high population. The population of the $i$-th 
type of cells at time $t+1$ is given by the dynamical map \cite{weiat88}
\begin{equation}
S_i(t+1) = \theta(h_i - \mu_i) \quad (i = 1,2,...,n) 
\end{equation}
where $\theta(y)$ is the step function, i.e., $\theta(y) = 0$ for $y < 0$ 
and $\theta(y) = 1$ for $y \geq 0$. Moreover, $h_i = \sum_j C_{ij} S_j$ 
is the total stimulus received by the cells of $i$-th type, where 
$C_{ij}$ is the interaction from cell type $j$ to the cell type $i$ and 
$\mu_i$ is a preassigned threshold at which $S_i$ switches from the state 
$0$ to the state $1$. The interactions $C_{ij}$ are allowed to take only 
a few integer values, for example, $-1, 0, 1$. A specific model is defined 
by the number $n$ of cell types, the set of values $\{C_{ij}\}$ for 
all the pairs $<ij>$ and the set of values $\{\mu_i\}$ for all $i$. If 
$S_i(t+1) = S_i(t)$ for all $i$ simultaneously then the corresponding 
values of the set $\{S_i\}$ is called a fixed point of the dynamics of 
the system. On the other hand, if $S_i(t+T) = S_i(t)$ for all $i$ 
simultaneously, then the system is said to have a limit cycle of period $T$. 
The fixed points and the limit cycles are referred to as the attractors 
of the dynamics. 

The concept of {\it window automata} has also been used extensively in 
theoretical immunology\cite{neuweis1, neuweis2}. Suppose, $S_i = 1$ only 
if $h_i$ falls within a certain window between two thresholds, i.e., if 
$\mu' < h_i < \mu''$, and $S_i = 0$ if $h_i$ falls outside this window; 
then $S_i$ is an example of window automaton. 

\noindent$\bullet${\bf Boolean Automata:} A Boolean automaton is a logical 
variable which can be only either "true" or "false", usually denoted by 
$1$ and $0$, respectively. Therefore, one can describe the cell populations 
in the discrete models by Boolean variables. However, one cannot carry out 
the standard algebraic operations, e.g., addition, multiplication, etc., 
with the Boolean variables. Therefore, if the discrete model is to be 
formulated in terms of Boolean automata, the dynamical maps will involve 
logical operations, viz., $OR, AND, NOT$, etc. For example, if $A, B$ and
$C$ are Boolean automata, then\\ 
$$A = B. OR. C \quad is \quad 1 \quad if \quad either \quad B \quad or \quad C \quad or \quad both \quad are \quad 1 $$
$$A = B. AND. C \quad is \quad 1 \quad if \quad and \quad only \quad if \quad both \quad B \quad and \quad C \quad are \quad 1 $$
$$A = .NOT.B \quad is \quad 1 \quad if \quad B \quad is \quad 0 \quad and \quad vice-versa. $$
The attractors of the dynamics of a Boolean automata network can also be 
defined just as we did in the case of threshold automata networks. 

We now present an illustrative example of a discrete model of NIR, within 
the framework of the clonal selection scenario, formulated using the language 
of Boolean automata; this model will be referred to as the e-KUT model 
as a model of this type was first considered by Kaufman, Urbain and Thomas 
\cite{kut85} and extended later by Chowdhury and Stauffer \cite{chow90a, chow90b}
Suppose, $Ab, S, H, B$ and $Ag$ denote the populations of the antibodies, 
suppressor $T$-cells, helper $T$-cells, $B$-cells and the corresponding 
foreign antigen, respectively. In the e-KUT model the dynamical maps 
governing the population dynamics are given by 
$$ Ab(t+1) = Ag(t).AND.B(t).AND.H(t) $$
$$S(t+1) = H(t).OR.S(t) $$
$$H(t+1) = [Ag(t).AND.(.NOT.S(t))].OR.H(t) $$
$$B(t+1) = [Ag(t).OR.B(t)].AND.H(t) $$ 
$$ Ag(t+1) = Ag(t).AND.(.NOT.Ab(t))$$ 
It is straightforward to check that this model has five fixed points each 
of which has a bio-medical interpretation. For example, the fixed point 
corresponding to $Ab = S = H = B = Ag = 0$ is the {\it virgin} or 
{\it tolerant} state whereas the fixed point corresponding to 
$Ag = Ab = 0$, $H = S = B = 1$ is interpreted as the immunized state where 
high populations of the lymphocytes carry the memory of the earlier encounter 
with the foreign antigen to which this clone responds specifically. 
Although the separate existence of suppressor $T$-cells is questionable, 
in the original model of Kaufman et al. \cite{kut85} a different 
interpretation of the origin of this suppressing effect was proposed; 
however, that interpretation has been criticised by Hoffmann\cite{hoff87}.
Not only the fixed points of the e-KUT model have interesting bio-medical 
interpretations, but the sequence of the intermediate states, through which 
the system evolves from an initial state before reaching the corresponding 
fixed point, have also been found to be consistent with experimentally known 
facts. 

Chowdhury et al.\cite{chow90b} developed a "unified" model, by generalizing 
the e-KUT model, which also describes the population dynamics of the cells 
involved in NIR, AIR as well as NIR against non-HIV antigens in HIV-infected 
individuals and AIDS. Suppose, the concentrations of the "self-antigen" and 
HIV are denoted by the symbols $IO$ and $IV$, respectively, while the 
concentration of the killer or effector cells involved in the AIR is 
denoted by the symbol $IK$. The dynamical maps governing the population 
dynamics in this model are now postulated to be 
$$ Ab(t+1) = Ag(t).AND.B(t).AND.H(t) $$
$$S(t+1) = H(t).OR.S(t) $$
$$H(t+1) = [((Ag(t).OR.IO).AND.(.NOT.S(t)).OR.H(t)].AND.(.NOT.IV) $$
$$B(t+1) = [Ag(t).OR.B(t)].AND.H(t) $$ 
$$ Ag(t+1) = Ag(t).AND.(.NOT.Ab(t))$$ 
$$ IK(t+1) = (IO(t).AND.H(t)).AND.(.NOT.S)$$
Cellular-automata models for some other aspects of immune response 
have also been developed \cite{pandey1, pandey2, celsei}.

\subsection{Continuum models of clonal selection} 

Starting from the discrete dynamical equations, written in terms of 
logical operations among boolean variables, it is possible to derive an  
analogous system of differential equations of the form 
\begin{equation}
(dy_i/dt) = g_i(y_1,y_2,...,y_N) - d_i y_i 
\end{equation}
for the $n$ types of cells where $y_i$ and $d_i$ represent, respectively, 
the concentration and the natural decay rate of the $i$-th type of cell. 
The functions $g_i$ involve combinations of sigmoid functions of $y_j$'s. 
Their functional forms can be derived from the form of the right hand side 
of the corresponding maps in the discrete theories following a well-defined 
prescription \cite{kut85,kut87}. In order to derive the right-hand side of the 
differential equation for the concentration $y_i$ from the right-hand side 
of the corresponding discrete map for the discrete variable $S_i$: (i) each 
of the discrete variables $S_j$ is replaced by the corresponding sigmoid 
function 
$$ F_i^+(y_j) = y_j^m/(\theta_{ij}^m + y_j^m) $$ 
whereas each of the discrete variables $.NOT.S_j$ is replaced by the 
function 
$$ F_i^-(y_j) = 1 - F_i^+(y_j) = \theta_{ij}^m/(\theta_{ij}^m + y_j^m) $$ 
where the Hill number $m$ determines the steepness of the sigmoid functions 
$F^+$ and $F^-$ and $\theta_{ij}$ is the threshold for the regulation of 
the cell type $i$ by the cell type $j$; (ii) the logical operations $OR$ 
and $AND$ used in the discrete formulation are replaced by the arithmetic 
operations of addition $(+)$ and multiplication $(\bullet)$, respectively, 
in the continuum formulation; (iii) an additional term of the form 
$- d_i y_i$ (with $d_i > 0$) is introduced to account for the natural decay 
of the populations of the cells of type $i$ with the passage of time. For 
example, the differential equation corresponding to the discrete equation 
$$S_3 = (S_5.AND.(.NOT.S_2)).OR.S_3 $$ 
is given by 
$$ (dy_3/dt) = k_3 F_3^+(y_5) \bullet F_3^-(y_2) + k_3' F_3^+(y_3) - d_3 y_3.$$

Following the prescriptions outlined above, Chowdhury \cite{chow93} derived 
the differential equations corresponding to the discrete dynamical maps in 
the unified model and further simplified the differential equations. 

An interesting feature of the NIR in this model is shown in fig.2. Note that 
a small amount of the antigen is adequate to immunize the host so that 
it can mount a very strong secondary response against the same antigen even 
if the antigen dose is high; this captures the essential principle of 
immunization or vaccination. 

Another interesting feature of this model is demonstrated in fig.3. If the 
host is infected with HIV but the concentration of HIV is low a secondary 
response to non-HIV antigens can take place despite depletion of the (memory-) 
$T_H$-cell populations. On the other hand, no secondary response to non-HIV 
antigen takes place if the concentration of HIV is sufficiently high. Thus, 
symptoms of AIDS (namely, lack of response to secondary antigens) would 
not be visible in an individual already infected by HIV, provided the level 
of HIV is low.

\section{Beyond Clonal Selection; Immune Network:} 

Clonal selection theory has been very successful in describing many aspects 
of immune response, but some crucial questions could not be answered so far 
within the framework of this theory. For example, what makes the memory 
cells retain their memory? One possibility is that some kind of stimulation 
of the immune system persists even after the antigen population falls below 
the tolerance level; in that case memory cells are nothing but cells which 
are perpetually in a stimulated state. But, if so, what keeps stimulating 
these cells so selectively and how \cite {vite91, sprent94}? Some 
experiments indicate that persistence of some traces of the foreign antigen 
after primary response can stimulate the "memory" $T-$ and $B-$cells 
\cite{gray88,fish95}. But, although this mechanism may be sufficient, 
this may not always be necessary as demonstrated by more recent experiments 
\cite{lau94, hou94}. 

A possible clue to this mystery of the identity of the specific stimulators, 
which keeps stimulating a clone so selectively long after the elimination 
of the foreign antigens, emerges from other sets of experiments. The clonal 
selection theory, in its classical form, assumed that all immune responses are 
triggered by antigens. But, it has been observed that in "germ-free" mice 
(i.e, mice kept for a few generations in environments free from foreign 
antigens) the number of activated lymphocytes is similar to the values 
measured in conventionally raised mice \cite{hoy84, perera86,varcout91} 
This observation suggests the possibility of 
stimulating clones also through internal mechanisms. We shall now argue that 
such internal mechanisms of stimulation follow naturally by going beyond the 
classical clonal selection theory and invoking the concept of an immune 
network. This network theory may also explain more satisfactorily some other 
immunological phenomena, e.g., tolerance and self-nonself discrimination, etc. 

Consider two clones $C_1$ and $C_2$. Suppose, the surface receptors of the 
lymphocytes and the free antibodies belonging to $C_1$ and $C_2$ "fit" with 
the epitopes of the foreign antigens $Ag_1$ and $Ag_2$, respectively. 
Therefore, according to the clonal selection theory, $C_1$ is expected to 
respond specifically to $Ag_1$ whereas $Ag_2$ is expected to stimulate $C_2$ 
selectively. It is quite natural to expect that the "molecular pattern" 
of the receptor molecules, which can recognize "molecular pattern" stored 
in the epitopes of foreign antigens, can themselves be recognized by others. 
For example, if the surface receptors of $C_1$ and $C_2$ "fit" with each 
other then $C_2$ would response to the proliferating lymphocytes of $C_1$ 
in exactly the same manner in which it responds against $Ag_2$. In other 
words, $C_2$ treats $C_1$ and $Ag_2$ on the same footing; therefore, $C_1$ 
may be regarded as an "internal image" of $Ag_2$. An epitope that is unique 
to the surface receptors and antibodies of a specific type is called an 
idiotope. Hence, a functional network formed on the basis of idiotope 
recognition is usually referred to as idiotypic network \cite{jerne73, 
jerne74, jerne84}.

Proponents of the immune network theory \cite{perel89, atcohen89, stewvar89, 
cout89, varcout91, boeretal92a} maintain that the immune response to foreign 
antigens is offered by the entire immune system (or, at least, more than 
one clone) in a collective manner although the dominant role may be played 
by a single clone whose cell surface receptors "fit" best with the epitope 
of the specific invading antigen. However, the proliferating cells and 
antibodies of the responding clone (idiotype) trigger the response of the 
corresponding anti-idiotypes which, in turn, can stimulate their own 
anti-idiotypes, and so on. The detailed dynamics of the immune network, of 
course, would depend on the size and the nature of the connectivity. 

$\bullet${\it Linear and cyclic networks:}  Richter \cite{richter75} introduced 
the earliest models of immune networks where a "chain-reaction" of the 
clones was postulated.  Suppose, the clones are such that $C_1$ 
stimulates $C_2$, then $C_2$ stimulates $C_3$ which, in turn, stimulates 
$C_4$, etc. However, this chain reaction is limited by the fact that 
each clone suppresses the particular clone that was responsible for its 
stimulation, i.e., simultaneously, suppose, $C_4$ responds to suppress $C_3$, 
$C_3$ suppresses $C_2$ which, in turn, suppresses $C_1$. Hiernaux \cite{hier}
converted the linear chain into a cyclic network and analyzed its properties.
Hoffmann and coworkers \cite{hoff75, hoff88, royer94} have made several 
improvements over the Richter model. Farmer et al.\cite{farm86, farm87} 
introduced a similar model and compared its features with the other networks 
used in adaptive computation \cite{farm90}. This work has been subsequently 
extended \cite{behn89, behn93} by incorporating memory B-cells. The attractors 
of the dynamics of such networks can be a limit cycle, where the populations 
of the antibodies vary periodically and the immunological memory is stored 
through a combination of "static" elements (namely, long-lived memory 
cells) and a "dynamic" process, namely, a limit cycle. A discrete toy 
model of a cyclic immune network has been developed by Chowdhury et al.
\cite{chow94} and its continuum counterpart has been investigated 
\cite{chow95}.

$\bullet${\it Cayley-tree-like network:} A Cayley tree is a loop-less tree 
characterized by the coordination number $z$ which is the number of branches 
emerging from each node. In such a network, the clones are organized in a 
hierarchical manner; the total stimulus received by the $B$-cells 
belonging to each of the clones at the $i$-th level is \cite{weis90} 
\begin{equation}
h^i = x^{(i-1)} + (z-1) x^{(i+1)}  
\end{equation}
where $x$ denotes the concentration of the $B$-cells. The concentration 
of the $B$-cells of the clones at the $i$-th level are, then, assumed to 
be governed by window automata (in the discrete formulation) or their  
continuum counterparts (in the continuum formulation). 

$\bullet${\it Generalized shape-space approach:} This formulation exploits 
the fact that the binding affinity of the surface receptors and free 
antibodies belonging to different clones is determined by the degree of 
complimentarity of their geometric shape, electric charge, etc. Therefore, 
if the generalized shape (includes $d$ different characteristics) of a 
clone is represented by a lattice site at $\vec r$ in a $d$-dimensional space 
then the location of its anti-idiotype should be $-\vec r$ on the same $d$-
dimensional lattice (see fig.4). Thus the strength of the interaction between 
the clones at the lattice site $\vec r$ and $-\vec r$ is maximum. Moreover, 
since the clone at $\vec r$ has still significant amount of complimentarity 
in generalized shape with the nearest-neighbour sites of $-\vec r$ the clone 
at the site $\vec r$ is usually assumed to interact also with those at the 
sites $-\vec r \pm \vec \delta_x$ and $-\vec r \pm \vec \delta_y$, although 
these interactions are much weaker than that between $\vec r$ and $-\vec r$ 
\cite{perost79, segper89, bosepe, boeretal92a, stweis92, sahist93}. It is 
desirable  that if the virgin system is infected by a single specific antigen 
that the response activities should remain confined over a limited region of 
the generalized shape space in spite of the fact that the entire network is 
connected. After all, when a person gets infected by tuberculosis he is not 
expected to show large populations of antibodies against cholera! It has been 
found that the dimensionality of the generalized shape space determines 
whether the response activities remain localized or percolate over the entire 
network. 

Chowdhury et al. \cite{chow94} have extended the e-KUT model \cite 
{chow90a, chow90b} of clonal selection so as to incorporate explicitly both 
intra-clonal and inter-clonal interactions. They postulated that the 
immune system in a host consists of several functional networks of various 
different sizes; clones belonging to different networks do not interact 
among themselves and only clones belonging to the same network can interact 
among themselves through inter-clonal interactions.  For example, on a square 
lattice the dynamical maps governing the time evolution of the populations 
of the various cell types are given by 

$$Ab(\vec r,t+1) = Ag(\vec r,t).AND.B(\vec r,t).AND.H(\vec r,t) $$
$$B(\vec r,t+1) = [Ag(\vec r,t).OR.H(-\vec r,t)
                               .OR.H(-\vec r+\vec \delta_1,t)
                               .OR.H(-\vec r+\vec \delta_2,t)$$ 
$$                             .OR.H(-\vec r+\vec \delta_3,t)
                               .OR.H(-\vec r+\vec \delta_4,t)].AND.H(\vec r,t)$$
$$H(\vec r,t+1) = [Ag(\vec r,t).OR.H(-\vec r,t)
                               .OR.H(-\vec r+\vec \delta_1,t)
                               .OR.H(-\vec r+\vec \delta_2,t) $$
$$                               .OR.H(-\vec r+\vec \delta_3,t)
                               .OR.H(-\vec r+\vec \delta_4,t)]
							   .AND.[.NOT.[H(-\vec r,t)
                               .AND.H(-\vec r+\vec \delta_1,t)$$
$$                               .AND.H(-\vec r+\vec \delta_2,t)
                               .AND.H(-\vec r+\vec \delta_3,t)
                               .AND.H(-\vec r+\vec \delta_4,t)]]$$
$$Ag(\vec r,t+1) = Ag(\vec r,t).AND.[.NOT.Ab(\vec r,t)]$$
where $\vec r + \vec \delta_1, \vec r + \vec \delta_2, \vec r + \vec \delta_3, 
\vec r + \vec \delta_4$ denote the positions of the four nearest-neighbours 
of the site $\vec r$. Thus, every clone at $\vec r$ can stimulate not only 
the antiidiotype at $-\vec r$ but also those clones which are located at 
the nearest-neighbour sites of $-\vec r$. On the other hand, so far as 
suppression is concerned, the high population of the clone at $\vec r$ can 
be reduced to a low level if and only if the populations of the clone at 
$-\vec r$ and the clones at the latter's $2d$ neighbouring sites are high 
simultaneously. Therefore, on the one hand, mutual stimulation is symmetric 
in the sense that the clone at $\vec r$ excites the clone at $-\vec r$ and 
vice versa. On the other hand, mutual suppression is also symmetric in the 
sense that clones at $\vec r$ and its $2d$ neighbouring sites together can 
reduce the population of the clone at$-\vec r$ and, similarly, the clones 
at $-\vec r$ and its nearest-neighbours can reduce the high population of 
the clone at $\vec r$ to a low level. However, there is asymmetry between 
stimulation and suppression because suppression succeeds only if the entire 
neighbourhood of the antiidiotype, rather than the antiidiotype alone, is 
highly populated.  Chowdhury et al.\cite{chow94} observed that, in this 
model, once a site of the shape space is infected a pulse propagates and 
the pattern of the pulse keeps recurring for ever, thereby carrying the 
memory of the encounter with the foreign antigen through a dynamic mechanism. 

\section{Summary and Conclusion:} 

The immune system is an example of complex adaptive systems. Other 
important adaptive systems include the brain (neural network) \cite{amit89}. 
There are several striking similarities between the brain and the the 
immune system despite many crucial differences.
Adaptive systems learn or adapt as living systems do. Different systems 
learn on widely different time scales; for example, brains learn 
in seconds to hours, immune systems in hours to days, species in 
days to centuries and ecosystems in centuries to millenia.
In my opinion, there are at least two different aspects of the dynamics 
of the immune system: (a) the populations of the cells of a specific clone 
(and, perhaps, closely related clones) increase very rapidly following the 
recognition of any foreign antigen and, after the elimination of the antigen, 
decrease again; (b) because of the natural death of unstimulated lymphocytes 
and recruitment of fresh immunocompetent cells the immune network itself 
evolves with time and all its characteristic properties, e.g., size, 
connectivity, etc., may also keep changing with time. Both the processes 
(a) and (b) occur on comparable time scales. Therefore, inclusion of 
both these aspects in the same model is more desirable than 
studying the cell-population dynamics in a network of fixed size and 
connectivity. In fact, not only the later evolution but also the formation 
of the immune network in a newborn child is a challenging problem. Even the 
size and the connectivity of the immune system may be among its emergent 
collective properties \cite{boerper91}. It has been speculated \cite{parisi88} 
that immunologcal memory would be much more robust if it is distributed 
over many clones rather than a single one. 

In this chapter we have not only explained some interesting methods 
of modelling in theoretical immunology but also presented some models 
as illustrative examples. Even if some of these models turn out to be 
inadequate to capture the complexities of the real immune system they 
may, nevertheless, find use in designing artificial immune systems 
for the protection of information systems (e.g., computer, internet, etc.) 
against the corresponding "antigens" (e.g., computer virus and internet 
worms, etc.). 

{\bf Acknowledgements}: I thank D. Stauffer for enjoyable collaborations 
over almost a decade on the modelling of immune networks and for his comments 
on an earlier version of the manuscript.

\newpage
{\bf Figure Captions:} 

{\bf Fig.1}: A schematic description of the routes to immunity. 

{\bf Fig.2}: The time-dependence of the population of the antibodies during 
the primary and secondary NIR for different antigen-dosages. The primary and 
secondary doses are given at $t = 0$ and $t = 100$, respectively. The 
strengths of the primary and secondary doses of antigen are both $1$ (in (a)), 
$10$ (in (b)), $100$ (in (c)) whereas those of the antigen are, respectively, 
$1$ and $100$ (in (d)).

{\bf Fig.3}: The time-dependence of the helper-T cells (full line), 
suppressor-T cells (dashed line), the antibodies (dotted line) and the antigen 
(asterisk-marked line) of a host, which has been immunized first against a 
specific non-HIV antigen, during an infection by different levels of HIV dose 
and the subsequent secondary response. The HIV dose is given at $t = 0$ and 
the secondary dose of the non-HIV antigen is given at $t = 150$ where the 
strength of the secondary dose of the non-HIV antigen is $5$. The levels of 
HIV doses are $0.5$ (in (a)), $1.5$ (in (b)) and $5.0$ (in (c)). 

{\bf Fig.4}: A schematic representation of the two-dimensional generalized 
shape space.  


\newpage

\begin{thebibliography}{99} 

\bibitem{silver89} Silverstein A M, 1989, {\it A history of immunology} (Academic Press) 

\bibitem{albert89}Alberts et al., 1989, {\it The Molecular Biology of the Cell}, 2nd ed. (Garland, New York) 

\bibitem{darnel90}Darnell et al., 1990, {\it Molecular Cell Biology} (Freeman/Scientific American, 
San Francisco) 

\bibitem{jaret86}Jaret P and Nilsson L, 1986, National Geographic (June) 702 

\bibitem{perweis}Perelson A S and Weisbuch, 1997, Reviews of Modern Physics {\bf 69}, 1219

\bibitem{chow92}Chowdhury D and Stauffer D, 1992, Physica A {\bf 186}, 61 

\bibitem{perel88}Perelson A S, 1988, (ed.) {\it Theoretical Immunology} parts I and II (Addison-Wesley) 

\bibitem{marchuk83}Marchuk G I, 1983, {\it Mathematical Models in Immunology}, (Optimization 
Software Inc., New York) 

\bibitem{burnet57}Burnet F M, 1957, Australian Journal of Science {\bf 20}, 67; see more detailed account in {\it The Clonal Selection Theory of Acquired Immunity} (Cambridge University Press, 1959); see also Scientific American {\bf 204(1)}, 58 (1961) 

\bibitem{reni90}Rennie J, 1990, Scientific American (12) 77; see also the special issue on "Frontiers in Biotechnology: Tolerance in the immune system" of Science, {\bf 248} (15th June) 1990. 

\bibitem{cohen88} Cohen I R, 1988, Scientific American {\bf 258}, 34; see also {\it Perspectives on Autoimmunity} (CRC press, Boca Raton, 1988) 

\bibitem{weber}Weber J N and Weiss R A, 1988, Scientific American (1) 81; see also the other articles in this special issue on AIDS 

\bibitem{bell70}Bell G I, 1970, Journal of Theoretical Biology {\bf 29}, 191 

\bibitem{bell71}Bell G I, 1971, Journal of Theoretical Biology {\bf 33}, 339  

\bibitem{atlan89}Atlan H, 1989, Bulletin of Mathematical Biology {\bf 51}, 247 

\bibitem{wolf86}Wolfram S, 1986, {\it Theory and Applications of Cellular Automata} (World Scientific) 

\bibitem{weisbook91}Weisbuch G, 1991, {\it Complex Systems Dynamics: An Introduction to Automata Networks} (Addison-Wesley) 

\bibitem{weiat88}Weisbuch G and Atlan H, 1988, J. Phys. A {\bf 21}, L189; see also  
      Cohen I R and Atlan H, 1989, J. Autoimmunity, {\bf 2}, 613 

\bibitem{neuweis1} Neumann A U and Weisbuch G, 1992, Bulletin of Mathematical Biology. {\bf 54}, 21 

\bibitem{neuweis2} Neumann A U and Weisbuch G, 1992, Bulletin of Mathematical Biology. {\bf 54}, 699 

\bibitem{kut85}Kaufman M, Urbain J and Thomas R, 1985, Journal of Theoretical Biology {\bf 114}, 527 

\bibitem{chow90a}Chowdhury D and Stauffer D, 1990, Journal of Statistical Physics {\bf 59}, 1019 

\bibitem{chow90b}Chowdhury D, Stauffer D and Choudary P V, 1990, Journal of Theoretical Biology {\bf 145}, 207 

\bibitem{hoff87}Hoffmann G W, 1987, Journal of Theoretical Biology {\bf 129}, 355 

\bibitem{pandey1} Pandey R B, 1991, Physica A {\bf 179}, 442  

\bibitem{pandey2} Pandey R B and Stauffer D, 1990, J. Stat. Phys. {\bf 61}, 235 

\bibitem{celsei} Celada F and Seiden P E, 1996, Eur. J. Immunol. {\bf 26}, 1350 


\bibitem{kut87}Kaufman M and Thomas R, 1987, Journal of Theoretical Biology {\bf 129}, 141 

\bibitem{chow93}Chowdhury D, 1993, Journal of Theoretical Biology {\bf 165}, 135 

\bibitem{vite91} Vitetta E S, Berton M T, Burger C, Kepron M, Lee W T and Yin Xiao-Ming, 1991, Annual Reviews of Immunology {\bf 9}, 193 

\bibitem{sprent94} Sprent J, 1994, Cell {\bf 76}, 315 

\bibitem{gray88} Gray D and Skarvall H, 1988, Nature {\bf 336}, 70 

\bibitem{fish95} M.A. Fishman and A.S. Perelson, 1995, Journal of Theoretical Biology {\bf 173}, 241 

\bibitem{lau94}Lau L L, Jamieson B D, Somasundaram T and Ahmed R, 1994, Nature {\bf 369}, 648 

\bibitem{hou94}Hou S, Hyland L, Ryan K W, Portner A and Doherty P C, 1994, Nature {\bf 369}, 652 

\bibitem{hoy84}Hoykaas H, Benner R, Pleasants R and Wostmann B, 1984, European Journal of Immunology {\bf 14}, 1 

\bibitem{perera86} Pereira P, Forni L, Larsson E L, Cooper M, Heisser C and Coutinho A, 1986, European Journal of Immunology {\bf 16}, 685 

\bibitem{jerne73}Jerne N K, 1973, Scientific American {\bf 229}, 52 

\bibitem{jerne74}Jerne N K, 1974, Ann. Immunol. (Inst. Pasteur) {\bf 125C}, 373 

\bibitem{jerne84}Jerne N K, 1984, Immunological Reviews {\bf 79}, 5 

\bibitem{perel89}Perelson A S, 1989, Immunological Reviews {\bf 110}, 5 

\bibitem{atcohen89}Atlan H and Cohen I R, 1989, (eds.) {\it Theories of Immune Networks} (Springer) 

\bibitem{stewvar89}Stewart J and Varela F J, 1989, Immunological Reviews {\bf 110}, 37 

\bibitem{cout89}Coutinho A, 1989, Immunological Reviews {\bf 110}, 63 

\bibitem{varcout91}Varela F J and Coutinho A, 1991, Immunology Today {\bf 12}, 159 
\bibitem{boeretal92a}De Boer R J, Neumann A U, Perelson A S, Segel L A and Weisbuch G, 1992a, in: Proc. of the first European Biomathematics Conference, eds. V. Capasso and P. Demongot (Springer) 

\bibitem{richter75}Richter P H, 1975, European Journal of Immunology {\bf 5}, 350 

\bibitem{hier} Hiernaux J, 1977, Immunochemistry, {\bf 14}, 733 

\bibitem{hoff75}Hoffmann G W, 1975, European Journal Immunology {\bf 5}, 638 

\bibitem{hoff88}Hoffmann G W, Kion T A, Forsyth R B, Soga K G and Cooper-Willis,1988, in: {\it Theoretical Immunology}, part Two, ed. Perelson A S (Addison-Wesley) 

\bibitem{royer94} S. Royer, 1994, Masters Thesis, Univ. of British Columbia. 

\bibitem{farm86} Farmer J D, Packard N H and Perelson A S, 1986, Physica D {\bf 22}, 187 

\bibitem{farm87} Farmer J D, Kaufman S A, Packard N H and Perelson A S, 1987, in: {\it Perspectives in Biological Dynamics and Theoretical Medicine}, eds. S.H. Koslow, A.J. 
Mandell and M.F. Schlesinger (New York Academy of Sciences, New York)

\bibitem{farm90}Farmer J D, 1990, Physica D {\bf 42}, 153 

\bibitem{behn89}Behn U and Van Hemmen J L, 1989, Journal of Statistical Physics {\bf 56}, 533 

\bibitem{behn93}Behn U, Van Hemmen J L and Sulzer B, 1993, Journal of Theoretical Biology {\bf 165}, 1; see also Lippert K and Van Hemmen J L, 1997, in: {\it Annual Reviews of Computational Physics}, ed. D. Stauffer, vol.V (Wold Scientific) 

\bibitem{chow94}Chowdhury D, Deshpande V and Stauffer D, 1994, International Journal of Modern Physics {\bf C5}, 1049 

\bibitem{chow95} Chowdhury D, 1995, Indian Journal of Physics, {\bf 69B}, 539  

\bibitem{weis90}Weisbuch G, De Boer R J and Perelson A S, 1990, Journal of Theoretical Biology {\bf 146}, 483 

\bibitem{perost79}Perelson A S and Oster G F, 1979, Journal of Theoretical Biology {\bf 81}, 645 

\bibitem{segper89} Segel L A and Perelson A S, 1989, in: {\it Theories of Immune Networks}, eds. Atlan H and Cohen I R (Springer) 

\bibitem{bosepe}De Boer R J, Segel L A and Perelson A S, 1992b, Journal of Theoretical Biology {\bf 155}, 295 

\bibitem{stweis92}Stauffer D and Weisbuch G, 1992, Physica A {\bf 180}, 42 

\bibitem{sahist93} Sahimi M and Stauffer D, 1993, Physical Review Letters {\bf 71}, 4271  

\bibitem{amit89}Amit D J, 1989, {\it Modelling Brain Functions} (Cambridge University Press) 

\bibitem{boerper91}De Boer R J and Perelson A S, 1991, Journal of Theoretical Biology {\bf 149}, 381 

\bibitem{parisi88}Parisi G, 1988, in: {\it Chaos and Complexity}, eds. R. Livi, S. Ruffo, S. Ciliberto and M. Buiatti (World Scientific) 

\end{thebibliography}
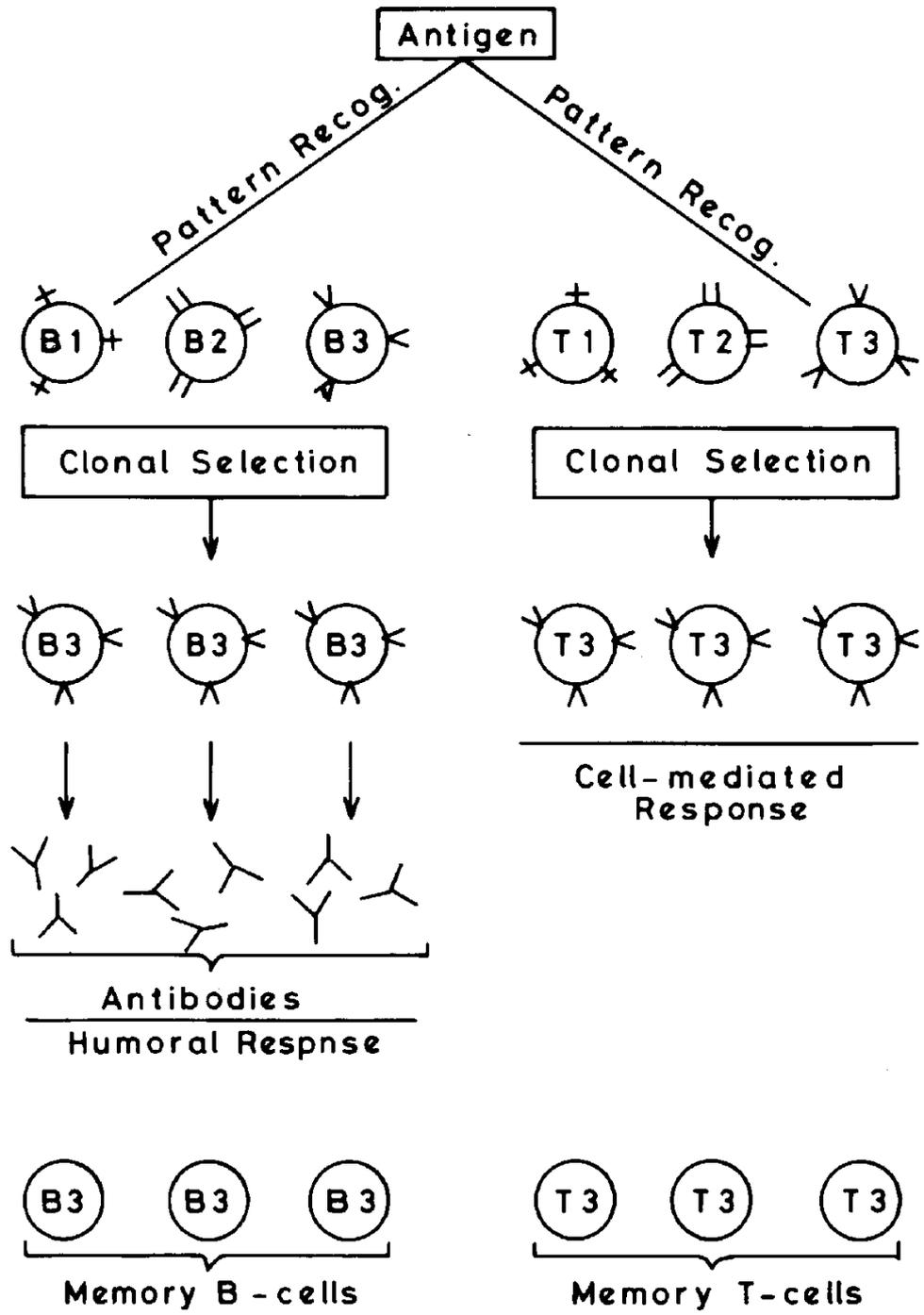
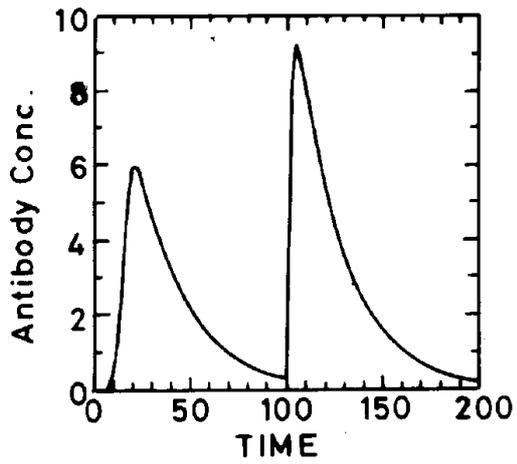
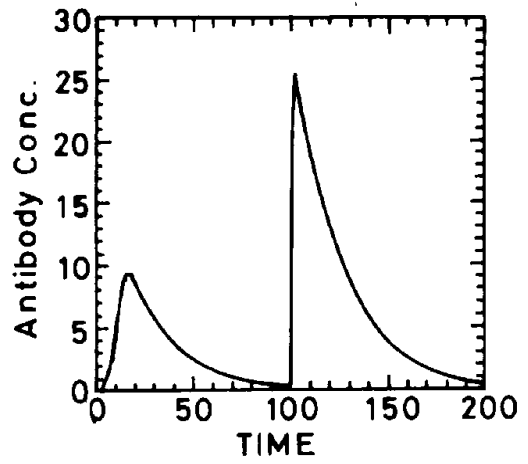
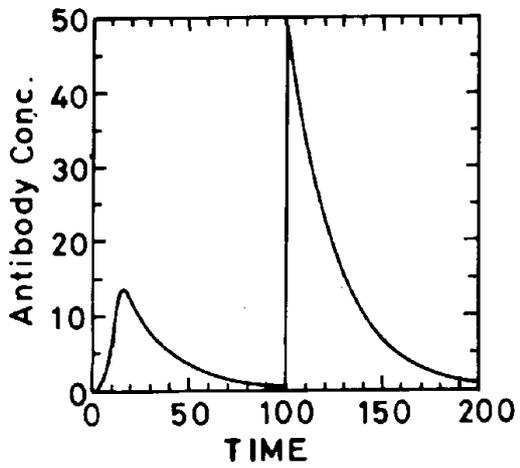
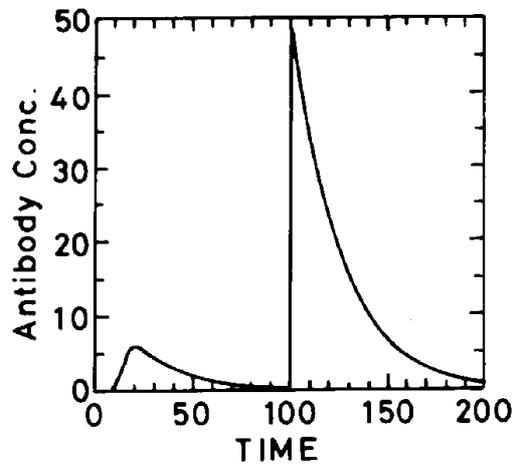
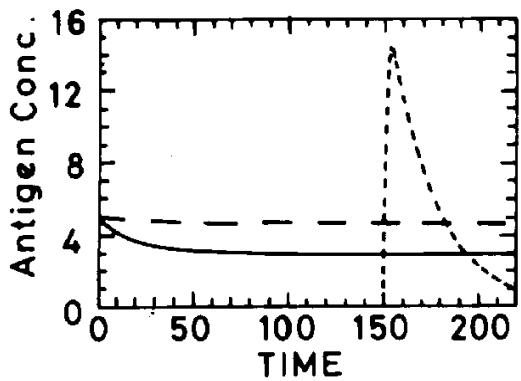
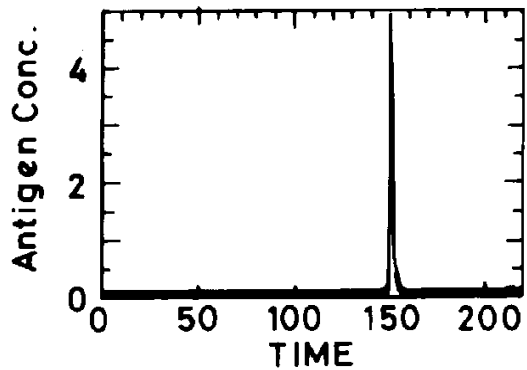
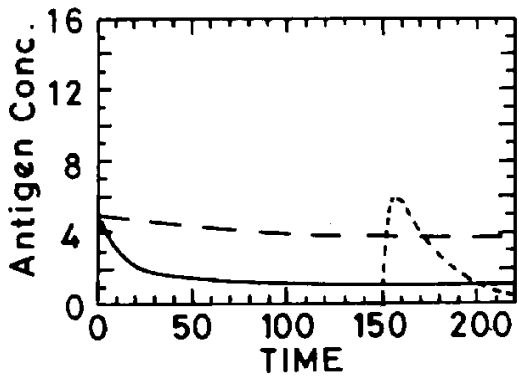
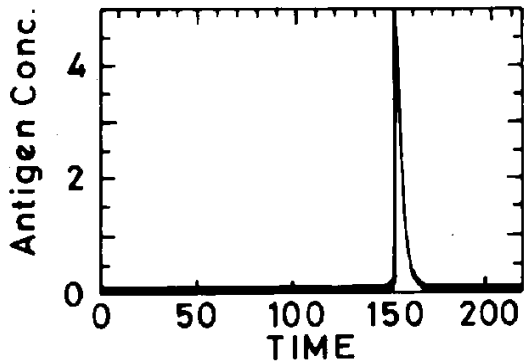
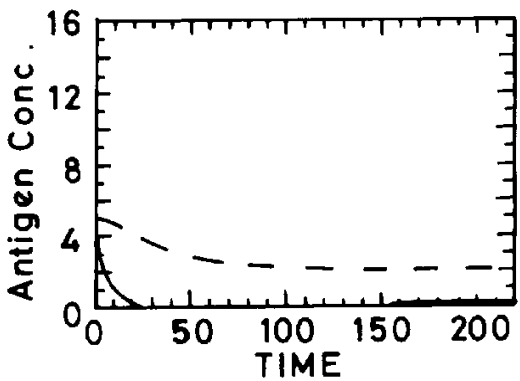
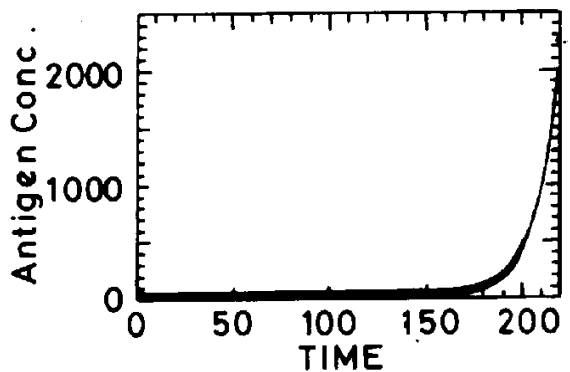
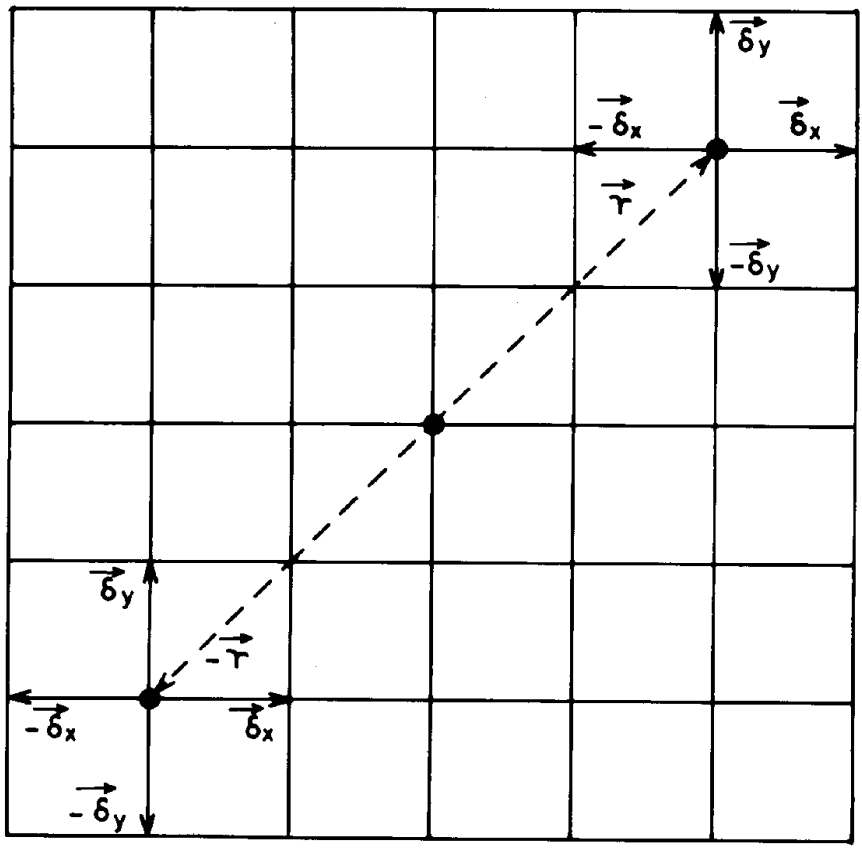
\end{document}